\documentclass[aip,reprint,a4paper] {revtex4-1}
\usepackage{upgreek}
\usepackage{amsfonts}
\usepackage{amstext}
\usepackage{mathtools}
\usepackage{amssymb}
\usepackage{color}
\usepackage[normalem]{ulem}
\usepackage{amsmath} 
\usepackage{graphicx} 
\usepackage{bbold}

\newcommand{\dir}{/home/gs1002/Bib/}

\newcommand{\tabClusterCoords}   {S1}

\newcommand{\figClusterCoords}   {S1}
\newcommand{\figCCtransitions}   {S2}
\newcommand{\figLigandContacts}  {S3}

\begin{document}

\title{Allosteric communication mediated by protein contact clusters:\\ A
  dynamical model}
\author{Ahmed A.\ A.\ I.\ Ali}
\altaffiliation{A.\ Ali and E.\ Dorbath contributed equally to this work.}
\author{Emanuel Dorbath}
\altaffiliation{A.\ Ali and E.\ Dorbath contributed equally to this work.}
\author{Gerhard Stock}
\email{stock@physik.uni-freiburg.de}
\affiliation{Biomolecular Dynamics, Institute of Physics,
   University of Freiburg, 79104 Freiburg, Germany}
\date{\today}

\begin{abstract}

  Allostery refers to the puzzling phenomenon of long-range
  communication between distant sites in proteins. Despite its
  importance in biomolecular regulation and signal transduction, the
  underlying dynamical process is not well understood. This study
  introduces a dynamical model of allosteric communication based on
  ``contact clusters"—localized groups of highly correlated contacts
  that facilitate interactions between secondary structures. The model
  shows that allostery involves a multi-step process with cooperative
  contact changes within clusters and communication between distant
  clusters mediated by rigid secondary structures.  Considering
  time-dependent experiments on a photoswitchable PDZ3 domain,
  extensive (in total $\sim 500\,\mu$s) molecular dynamics simulations
  are conducted that directly monitor the photoinduced allosteric
  transition. The structural reorganization is illustrated by the time
  evolution of the contact clusters and the ligand, which effects the
  nonlocal coupling between distant clusters. A timescale analysis
  reveals dynamics from nano- to microseconds, which are in excellent
  agreement with the experimentally measured timescales.

\end{abstract}
\maketitle

%
%

\section{Introduction}
\vspace*{-4mm}

The investigation of allostery, which accounts for the intriguing
phenomenon of long-range communication between distant protein sites,
has attracted considerable attention in both experimental and
computational domains. \cite{Gunasekaran04, Bahar07,Cui08, Changeux12,
  Motlagh14,Tsai14,Thirumalai19,Wodak19} Considering its critical role
in cellular signaling and as a target in pharmaceutical research,
however, our basic understanding of the dynamic process underlying
allostery remains surprisingly limited. The direct observation of
allosteric transitions remains difficult, particularly due to the
subtle nature of structural changes\cite{Brueschweiler09,
  Mehrabi19,Bozovic22} and sampling limitations of molecular dynamics
(MD) simulations. \cite{Mehdi24} Moreover, the few existing MD studies
showing allosteric transitions are found to be difficult to interpret,
because they typically involve the nonlinear and nonlocal response of
many protein residues. \cite{Vesper13,Pontiggia15,
  Buchenberg17,Zheng18,Ayaz23,Vossel23}
Most commonly, allosteric communication is explained by network
models,\cite{Guo16,Dokholyan16} where the protein residues are the
'nodes' of the network and the 'edges' describe some inter-residue
coupling. While these formulations have contributed considerably to
our understanding of allostery, they are by design statistical models,
rather than a dynamical model of the real-time evolution of an
allosteric transition. In this work, we aim to develop such a
dynamical model.

\begin{figure}[ht!]
	\centering
	\includegraphics[width=0.4\textwidth]{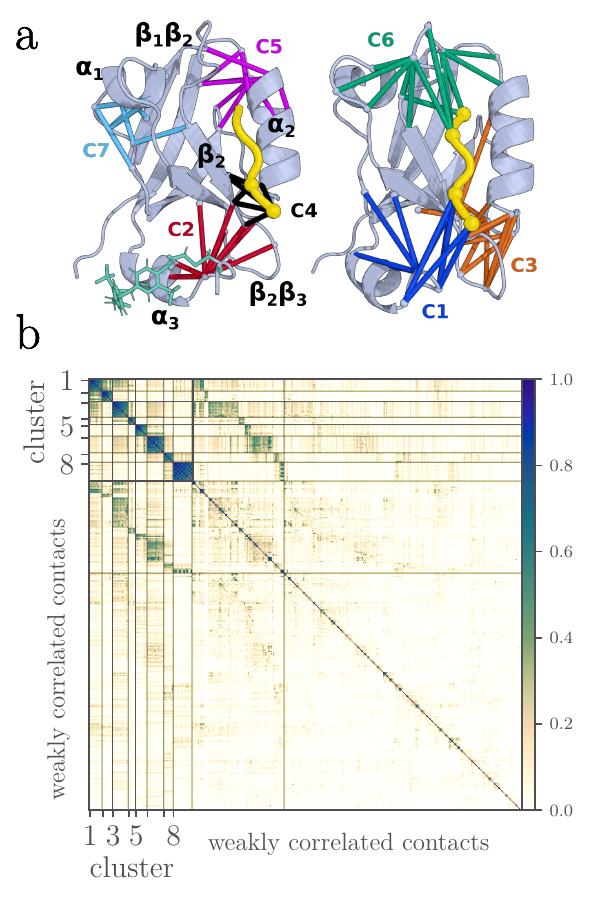}
	\caption{
          (a) Photoswitchable PDZ3 domain,\cite{Bozovic21} indicating
          main secondary structural elements, the ligand (KETWV,
          yellow), and the azobenzene photoswitch (green, only shown
          on the left side). Inserted colored lines illustrate the
          contact distances associated with clusters C1 to C7 obtained
          from MoSAIC.\cite{Diez22} (b) Block-diagonal correlation
          matrix obtained from MoSAIC clustering of all contact
          distances of PDZ3.}
	\label{fig:pdz}
\end{figure}

To gain an understanding of the dynamical process underlying
allostery, it is helpful to focus on a simple model system that allows
us to perform time-dependent experiments as well as MD simulations,
which both may directly observe the allosteric transition. In this
respect, PDZ domains have been extensively studied, because they are
well-established and structurally conserved protein interaction
modules involved in the regulation of multiple receptor-coupled signal
transduction processes, but at the same time have also been considered
as isolated model systems of allosteric
communication. \cite{Fuentes04,Petit09, Ye13,Gautier19,
  Gerek11,Kumawat17, Faure22} They share a common fold
(Fig.\ \ref{fig:pdz}a), which consists of two $\alpha$-helices and six
$\beta$-strands, with the second $\alpha$-helix and the second
$\beta$-strand forming the canonical ligand binding
groove. Interestingly, the NMR study conducted by Petit et
al.\cite{Petit09} demonstrated that the removal of an additional short
$\alpha$-helix at the C-terminal of PDZ3 reduces ligand affinity by a
factor of 21, thereby revealing allosteric communication between the
C-terminal and ligand binding. In this sense, PDZ3 can be considered
as one of the smallest allosteric proteins, because both the active
site (the $\alpha_3$-helix) and the allosteric site (the binding
pocket) are clearly defined.

Recently, Bozovic et al.\cite{Bozovic21} employed photoswitching of
the $\alpha_3$-helix to trigger a conformational change in PDZ3,
propagating from the $\alpha_3$-helix to the ligand-binding
pocket. Using time-resolved vibrational spectroscopy, they obtained a
timescale of 200~ns for the perturbation to traverse to the ligand,
which was interpreted as the speed of signaling in this single-domain
protein.
To facilitate a microscopic understanding of these experiments, Ali et
al.\cite{Ali22} conducted extensive MD simulations of the system. By
identifying all inter-residue contacts that changed during these
simulations, they found a network of contacts linking the
$\alpha_3$-helix and the protein core, which mediated the observed
conformational transition. Hence, the study provided a first idea of the
atomistic mechanism and also roughly reproduced the experimentally
observed timescales.

However, the transient infrared response of PDZ3 is much more complex
than the single 200\,ns component due to the ligand. In line with
previous studies on photoswitchable proteins,
\cite{Buchenberg17,Stock18,Bozovic20,Bozovic22} the experiments revealed
intricate signals across multiple timescales, from picoseconds to
milliseconds. These findings raise questions on the nature of the
structural dynamics underlying the timescales. What are these motions
and are they associated with some function of the protein? Do they
account for the allosteric communication from the initially excited
$\alpha_3$-helix to distant sites, even farther than the ligand? Are
these processes mediated by a network of inter-residue contacts as
well?

In this work, we show that the experimentally observed multiple
timescales can indeed be explained by long-distance communication
between localized ensembles (or clusters) of inter-residue
contacts. These clusters can be identified by a correlation analysis
\cite{Diez22} that discriminates collective motions underlying
functional dynamics from uncorrelated motion. While the contact
clusters represent the flexible joints and hinges of the
protein,\cite{Papaleo16} the long-range communication between these
contact clusters is mediated by rigid secondary
structures.\cite{Hermans17} For example, we find a structural
rearrangement of the $\beta_1$-$\beta_2$ loop, which is triggered by
the photoswitching of the $\alpha_3$-helix at the other end of the
protein (Fig.\ \ref{fig:pdz}a). This gives rise to a dynamical model
of allostery, where the initially induced structural strain on the
protein causes a multi-step structural reorganization process, which
consists of cooperative conformational transitions within a contact
cluster and of the sequential communication between clusters.

%
%
\section{Results and Discussion}
\vspace*{-4mm}

All results shown are based on nonequilibrium MD data of the
photswitchable PDZ3 domain used in experiment,\cite{Bozovic21}
combining $90 \!\times\! 1\, \mu$s-long and
$22 \!\times\! 10\, \mu$s-long trajectories (in total $\sim 1.5\times 10^6$
MD frames), see Methods.

\subsection{Definition of contact clusters}
\vspace*{-4mm}

The analysis of conformational dynamics requires the choice of input
coordinates or features.\cite{Ravindra20,Nagel23} In particular
inter-residue contact distances have been shown advantageous, as they
may directly account for tertiary interactions and scale linearly with
the size of the system. \cite{Ernst15} Because typically only a small
subset of those coordinates is involved in a specific biomolecular
process, we want to discard the remaining uncorrelated motions or
weakly correlated noise coordinates. To this end, we perform a MoSAIC
correlation analysis,\cite{Diez22} which allows us to distinguish
collective motions underlying functional dynamics and uncorrelated
motions. Hence it provides a versatile feature selection method, which
was successfully applied to identify relevant coordinates of the
folding of HP35 \cite{Nagel23} and of the functional motion of T4
lysozyme.\cite{Post22a} While Ali et al.\cite{Ali22} merely used
MoSAIC to identify the contact network between the photoexcited
$\alpha_3$-helix and the protein core, in this work we extend the
contact analysis to all regions of PDZ3.

In a first step, we identify all contacts between residues
$i$ and $j$ from the MD data using a distance criterion, see
Methods. We then calculate the linear correlation matrix between
contact distances $r_{ij}$ and  $r_{kl}$,
\begin{equation} \label{eq:Rhoij}
\rho_{ij,kl}=\frac{\langle \delta r_{ij} \delta r_{kl} \rangle}
{\langle \delta r_{ij}^2 \rangle^{1/2} \;\langle \delta r_{kl}^2
  \rangle^{1/2}}, 
\end{equation}
where $\delta r_{ij} = r_{ij} - \langle r_{ij}\rangle$ and the
brackets $\langle \ldots \rangle$ denote a combined ensemble and time
average over all MD data. To rearrange this matrix in an
approximately block-diagonal form, we employ a community detection
technique called Leiden clustering,\cite{Traag19} using the Python
package MoSAIC\cite{Diez22} with a resolution parameter
$\gamma \!=\! 0.5$.

The resulting block-diagonal correlation matrix yields eight main
blocks or clusters termed C1 to C8, see the upper left square of Fig.\
\ref{fig:pdz}b.
Within such a cluster, the coordinates are highly correlated (i.e., on
average $|\rho| \ge \gamma$), while the correlation between different
clusters is low (i.e., $|\rho| < \gamma$). Moreover, we have contacts
that are only weakly correlated (middle square) or hardly correlated
(lower right square) with other contacts, and can therefore be omitted
in the further analysis. This is, e.g., the case for stable contacts
and contacts on the protein surface that form and break frequently. In
the discussion below, we also disregard cluster C8, whose contacts
involve the N-terminus and are therefore highly fluctuating. 

Figure \ref{fig:pdz}a shows the contacts of clusters C1 to C7 inserted
into the structure of PDZ3, where the numbering is chosen to roughly
correspond to the spatial proximity of the clusters. Remarkably, we
find that almost all cluster contacts mediate tertiary interactions
between secondary structures. In this way, they represent the
flexible joints and hinges of the protein, \cite{Papaleo16} which
facilitate structural rearrangements. For example, the 11 contacts of
C1 connect the $\alpha_3$-helix and the core of the protein, and thus
control the alignment of the $\alpha_3$-helix to the protein. C2 also
involves the $\alpha_3$-helix, as it connects Phe100 of $\alpha_3$ to
the $\beta_2$-$\beta_3$ loop. C3 helps to stabilize the
$\beta_2$-$\beta_3$ loop through (up to) 14 interactions with
$\beta_4$-$\alpha_2$. The other main loop, $\beta_1$-$\beta_2$, is
stabilized via (up to) 16 contacts to the ligand and
$\beta_3$-$\alpha_1$ in C6. Moreover, C4 connects the ligand to
$\beta_2$, C5 bridges between $\alpha_2$ and $\beta_5$, and C7
connects $\beta_1$-$\beta_2$ and $\alpha_1$ to
$\alpha_1$-$\beta_4$. See Table \tabClusterCoords\ for a list of the
contacts of all clusters, which also demonstrates the good stability
of the clustering by comparing the outcome of different data sets.
In particular, it is shown that we obtain qualitatively the same
contact clusters from standard MD simulation of the {\em cis} and {\em
  trans} equilibrium states of the protein.

%
%
\subsection{Time evolution of contact clusters}
\vspace*{-4mm}

\begin{figure}[ht!]
	\centering
	\includegraphics[width=0.45\textwidth]{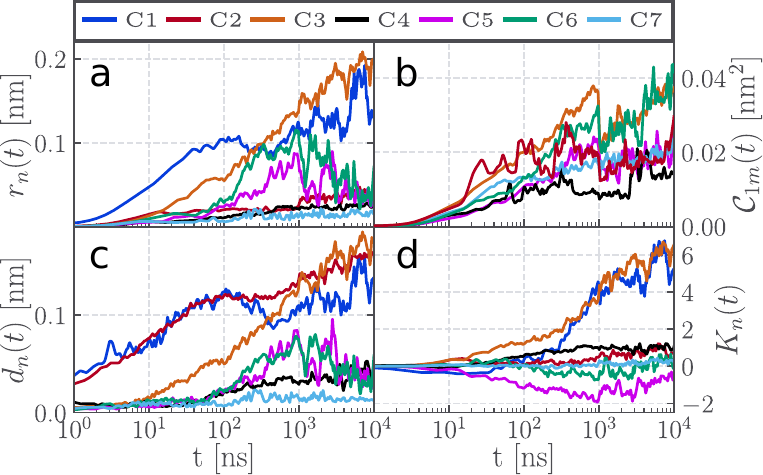}
	\caption{
          Time evolution of (a) the contact distances $r_n(t)$
          averaged over cluster $n$ [Eq.\ (\ref{eq:rkt})], (b) the
          intercluster correlation $C_{1m}(t)$ [Eq.\ (\ref{eq:Cnm})],
          (c) the corresponding cluster-averaged C$_\alpha$-distances
          $d_n(t)$, and (d) the average number of cluster contacts
          $K_n(t)$. We smoothed all log-time traces by a Gaussian
          filter function with a standard deviation of 6 frames. For
          $t \le 1 \mu$s, we average over 112 nonequilibrium MD
          trajectories, for $t > 1 \mu$s over only 22 trajectories,
          which explains the large fluctuations at long times.}
	\label{fig:clusters}
\end{figure}

Remarkably, the MoSAIC analysis achieves a partitioning of PDZ3 in
weakly interacting clusters of correlated contacts. We now wish to
study the nonequilibrium time evolution of these clusters, which 
follows the photoswitching of the $\alpha_3$-helix at time $t=0$ (see
Methods).
To this end, we introduce several observables that are averaged over all
contacts of a cluster. We begin with the average change of the contact
distances $r_{ij}$ in a cluster, defined as
\begin{align} \label{eq:rkt}
r_n(t)=\frac{1}{M_n}\sum_{i,j \in Cn} |\langle \Delta r_{ij}(t) \rangle_N| ,
\end{align}
where $\Delta r_{ij}(t) = r_{ij}(t) - r_{ij}(0)$, the brackets
$\langle \ldots \rangle_N$ denote the ensemble average over all $N$
nonequilibrium trajectories, and $M_n$ is the number of contact
distances in cluster~$n$. Note that we sum over the modulus of the
distance changes, in order to avoid canceling effects. Displaying the
time evolution of the cluster-averaged contact distances, Fig.\
\ref{fig:clusters}a as expected shows a rapid and strong response of
clusters C1 and the close-by C3. Interestingly, after about 100\,ns,
also the distant clusters C5 and C6 show a significant reaction. We
stress that the time-delayed response of the distant contacts is a
consequence of the initial photoswitching of cluster C1, and therefore
represents a direct evidence for allosteric coupling. (For an
equilibrium trajectory, we would find $r_n(t)\approx\,$const.)

We note that the finding of a significant response of the distant
clusters C5 and C6 is somewhat unexpected, because the corresponding
off-diagonal elements $\rho_{ij,kl}$ of the correlation matrix [Eq.\
(\ref{eq:Rhoij})] appear rather small in Fig.\ \ref{fig:pdz}b. To
explain this discrepancy, we define as alternative measure of the
correlation between clusters $n$ and $m$ the quantity
\begin{equation} \label{eq:Cnm}
  C_{nm}(t) = \frac{1}{M_nM_m} \sum_{i,j \in Cn} \sum_{k,l \in Cm}
  \!\!\langle | \Delta r_{ij}(t) \Delta r_{kl}(t) |\rangle_N .
\end{equation}
Displaying the time-dependent correlation $C_{1m}(t)$ of cluster C1
with the other clusters, Fig.\ \ref{fig:clusters}b clearly shows that
apart from the close-by cluster C3 also the distant cluster C6 gets
significantly excited due to the initial preparation of cluster
C1. Note that the overall rise of the intercluster correlations
$C_{nm}(t)$ directly reflects the time evolution of the associated
distances $\Delta r_{ij}(t)$ and $\Delta r_{kl}(t)$. Apart from the
missing time average and minor important normalization factors, the
main difference to Eq.\ (\ref{eq:Rhoij}) is that we take in Eq.\
(\ref{eq:Cnm}) the modulus of the correlation of each nonequilibrium
trajectory. This prevents the canceling of positive and negative
correlations of the individual trajectories.

Another quantity of interest is the average number of contacts
$K_n(t)$ of cluster $n$, see Fig.\ \ref{fig:clusters}d. We find that
only clusters C1 and the close-by C3 show a large change of $K_n(t)$
(both form about 6 contacts), while the other clusters change only
minor, $|K_n(t)-K_n(0)| \lesssim 1$. As an alternative distance
measure, Fig.\ \ref{fig:clusters}c finally shows the cluster-averaged
C$_\alpha$-distances, which are mostly quite similar to the contact
distances.

%
%
\subsection{Timescale analysis}
\vspace*{-4mm}

The quantities shown in Fig.\ \ref{fig:clusters} exhibit dynamics on
various timescales, ranging from nano- to microseconds. To provide a
well-defined measure for the timescales contained in such data, we
model the underlying time series $S(t)$ by a multiexponential response
function\cite{Stock18}
\begin{equation}\label{eq:Multiexp}
S(t) = \sum_{k} s_{k} \,e^{-t/\tau_{k}} .
\end{equation}
To this end, we choose the time constants $\tau_{k}$ to be equally
distributed on a logarithmic scale (e.g., 10 terms per decade) and fit
the corresponding amplitudes $s_k$ to the data, using a
maximum-entropy regularization method, \cite{Lorenz-Fonfria06} see
Methods.

\begin{figure}[ht!]
	\centering
	\includegraphics[width=0.3\textwidth]{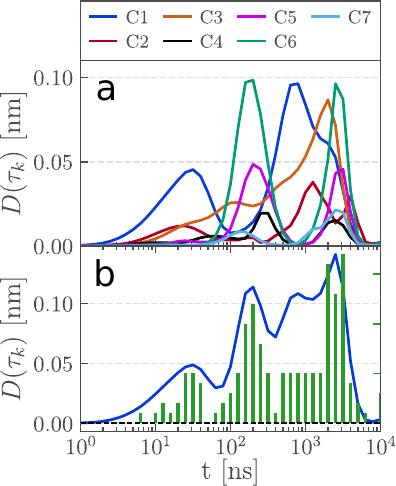}
	\caption{
          (a) Timescale spectrum of contact
          distances $r_{ij}(t)$, obtained for clusters C1 to C7. (b)
          Overall timescale spectrum calculated for all
          clusters. Green bars indicate the number of contact
          distances, whose timescale analysis shows a maximum at the
          considered timescale.}
	\label{fig:TA}
\end{figure}

To illustrate the distribution of timescales occurring in a cluster,
we perform the timescale analysis for each contact distance
$r_{ij}(t)$ contained in the cluster, and define the timescale
spectrum $D_n(\tau_k)$ of cluster $n$ as
\begin{equation}\label{eq:DynCont}
D_n(\tau_k) = \sqrt{ \sum_{i,j \in C_n}  |s_{k}(i,j)|^2 } ,
\end{equation}
which can be considered as the 'dynamical content' of the
cluster.\cite{Stock18} 
As shown in Fig.\ \ref{fig:TA}a, clusters C1, C3 and C6 show overall
the largest amplitudes, while the other clusters contribute
significantly weaker. Referring to the local maxima as the 'main
timescales' of the spectra, we find these timescales ranging from
30\,ns to $3\,\mu$s. Following photoswitching, clusters C1 and the
close-by C2 and C3 naturally exhibit the earliest response at
30\,ns. Less expected is that the second timescale at 200\,ns is
due to the distant clusters C6, C5 and C4 (in the order of their
importance). At 800\,ns we note the conformational transition of C1
aligning $\alpha_3$ to the protein (see below),
followed by transitions of the near-by clusters C2 and C3 at 1 and
1.1\,$\mu$s, respectively. Almost all clusters contribute to the
slowest timescale of $3\,\mu$s, with the largest contributions coming
from clusters C6, C3, and C5.

Apart from studying the response of each contact cluster separately,
we may also consider the response of the full system. To this end, we
perform a timescale analysis of the contact distances of all seven
clusters, and calculate the corresponding overall timescale spectrum
$D(\tau_k)$. (Formally the cluster-specific spectra $D_n(\tau_k)$ and
the overall spectrum $D(\tau_k)$ are related via
$D^2(\tau_k) = \sum_{n}  |D_n(\tau_k)|^2$.)
As shown in Fig.\ \ref{fig:TA}b, the overall timescale spectrum
exhibits well-defined peaks at 30\,ns, 200\,ns, and $3\,\mu$s, as well
as a broad shoulder at 800\,ns, which all can be readily associated with
the cluster-specific timescale spectra in Fig.\ \ref{fig:TA}a.

It is interesting to compare our calculated spectra to the
experimental results by Bozovic et al.,\cite{Bozovic21} who inferred
the timescale spectrum of PDZ3 from time-resolved infrared
spectroscopy. Remarkably, they found well-defined peaks at 20\,ns,
200\,ns, $2\,\mu$s and $20\,\mu$s, as well as a broad feature in the
decades following $\sim 1\,$ns and 200\,ns. Our analysis reproduces
perfectly the ligand-induced signal at 200\,ns, and matches well the
two intermediate times (30 vs.\ 20\,ns) and (3 vs.\ $2\,\mu$s). The
1\,ns timescale of the experimental spectrum can most likely be
explained by the initial 5\,ns stretching of the $\alpha_3$-helix (and
the associated breaking of the contacts $r_{28,102}$ and $r_{97,102}$
at $\lesssim 1\,$ns, see below); these features are only weakly
visible in Fig.\ \ref{fig:TA}. The diffuse region between the
experimental peaks at 200\,ns and $2\,\mu$s seems to contain several
contributions, which may be explained by the strongly overlapping
timescale spectra of clusters C1 and C3. Lastly, the long experimental
timescale of $20\,\mu$s is clearly out of reach for our study.


%
%
\subsection{Local response to photoswitching}
\vspace*{-4mm}

\begin{figure}[t!]
	\centering
	\includegraphics[width=0.4\textwidth]{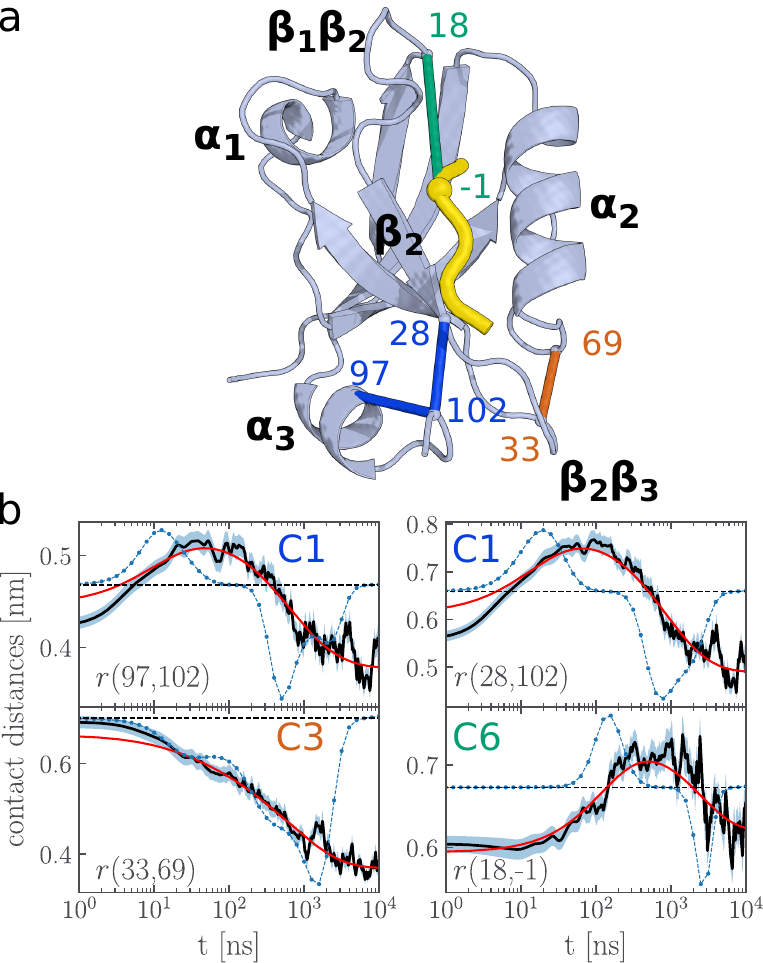}
	\caption{
          (a) Structural description and (b) time evolution of
          selected contact distances of clusters C1, C3, and C6. MD
          data are drawn in black, the timescale spectrum [Eq.\
          (\ref{eq:DynCont})] in blue, and the resulting fit of the
          data [Eq.\ (\ref{eq:Multiexp})] in red. Residues are
          numbered from 1 to 103 for the protein and from -4 to 0 for
          the ligand.}
	\label{fig:contacts}
\end{figure}

We are now in a position to develop a mechanistic picture of the
propagation of the initial perturbation of cluster C1 to the distant sites of
the protein. The discussion is based on selected contact distances of
the clusters shown in Fig.\ \ref{fig:contacts}, while Fig.\
\figClusterCoords{} provides comprehensive data of all these
distances.

%
Prior to photoswitching, the system is prepared in the {\em cis}
configuration, where the $\alpha_3$-helix is mostly stabilized by
nonpolar contacts with the protein (e.g., Val28 and Tyr97 connect to
Azo102). By attaching the azobenzene photoswitch to the side-chains of
residues 95 and 102 of the $\alpha_3$-helix, Bozovic et
al.\cite{Bozovic21} are able to switch the end-to-end distance of
azobenzene from its twisted {\em cis} configuration (accommodating a
stable helix) to the {\em trans} configuration (which destabilizes the
helix).
Following the sub-picosecond {\em cis}-to-{\em trans}
photoisomerization of azobenzene at time $t=0$, the photon energy is
to a large part converted to vibrational energy of the helix. Since
this kinetic energy is dissipated within tens of ps, it cannot disrupt
the stabilizing contacts. The transport of vibrational energy via the
backbone and the contacts of PDZ3, as well as the subsequent cooling
of the protein in the solvent water, was measured by Baumann et
al. \cite{Baumann19} and modeled in detail by Ishikura
et al.\cite{Ishikura15} and Gulzar et al.\cite{Gulzar19}

What is more, the photoswitch also introduces a local conformational
strain to the protein, which results in an
increase of the potential energy, and subsequently relaxes through a
sequence of processes on several timescales. In a first step, the
photoswitching stretches the $\alpha_3$-helix on a 5\,ns
timescale,\cite{Ali22} which affects the stabilizing contacts
$r_{97,102}$ and $r_{28,102}$ of cluster C1 (Fig.\
\ref{fig:contacts}). Defining the breaking of a contact by
$\langle r_{ij}(t) \rangle \gtrsim 0.45\,$nm, $r_{28,102}$ breaks at
$t=0.2\,$ns, and $r_{97,102}$ at 3\,ns. The resulting destabilization
of the $\alpha_3$-helix leads to the complete detachment of $\alpha_3$
from the protein, which occurs on a timescale on 30\,ns. This is
evident from the rise of several distances of C1 up to a maximum at
$t\sim 100\,$ns (Fig.\ \ref{fig:contacts}). Hence we have shown that
the peak at 30\,ns of the C1 timescale spectrum (Fig.\ \ref{fig:TA}a)
is caused by the photoinduced detachment of $\alpha_3$.

The other two timescales found for C1 are 800\,ns which reflects the
main conformational transition of C1, and a weaker feature at
$3\,\mu$s, which accounts for the structural relaxation of C1 towards
the equilibrium state of the {\em trans} configuration. Most notably,
the main conformational transition aligns $\alpha_3$ to the protein,
by cooperatively forming (at least) four new contacts, including three
salt bridges [(Azo102,Lys55), (Lys103,Glu(-3), (Lys103,Lys(-4))] and
one hydrogen bond (Glu101,Lys(-4)). \cite{Ali22} (Residues are
numbered from 1 to 103 for the protein and from -4 to 0 for the
ligand.)  This is evident from the decrease of all contact distances
for $t \gtrsim 100\,$ns, see e.g., $r_{97,102}$ and $r_{28,102}$ in
Fig.\ \ref{fig:contacts}. As shown by the mean number of contacts of
C1 (Fig.\ \ref{fig:clusters}), this process is only finished at
$t \sim 3\,\mu$s. While the previously used $1\,\mu$s-long
trajectories suggested a timescale of 300\,ns for the conformational
transition,\cite{Ali22} our $10\,\mu$s-long data in fact yield a
timescale of 800\,ns.
Considering the overall evolution from {\em cis} equilibrium
($t < 0$), photoexcitation ($t= 0$), conformational transition
($t \sim 1\,\mu$s), and relaxation towards {\em trans} equilibrium
($t \gtrsim 3\,\mu$s), cluster C1 describes an order-disorder-order
transition.\cite{Buchenberg17}

%
%
%
We turn to cluster C2, which is special as it only contains contacts
involving residue 100 of the $\alpha_3$-helix (Fig.\
\ref{fig:pdz}a). Contact partners are residues Glu96 and Tyr97 of
$\alpha_3$ and seven residues of the $\beta_2$-$\beta_3$-loop (from
Val28 to Gly35). The timescale spectrum of C2 shows two relatively
weak signatures, a peak at 30ns and a double-peak around $2.5\,\mu$s, which
originate from the contacts with $\alpha_3$ and $\beta_2\beta_3$,
respectively. Overall, though, C2 exhibits only minor changes of its
contact distances, and therefore acts mostly as a quite rigid
connection between the $\alpha_3$-helix of C1 and the
$\beta_2$-$\beta_3$-loop of C3.

Next to C1, cluster C3 is the other cluster that forms numerous new
contacts (up to six) during its time evolution, see Fig.\
\ref{fig:clusters}d. Consisting of 14 contact distances connecting the
$\beta_2$-$\beta_3$-loop (residues 29 - 36) and the
$\beta_4$-$\alpha_2$-loop (residues 67 - 71), this contact formation
leads to a stabilization of the flexible $\beta_2$-$\beta_3$-loop.  Hence
all distances are decreasing (e.g., $r_{33,69}$ in Fig.\
\ref{fig:contacts}), with the exception of $r_{29,71}$ reflecting the
widening of the binding pocket due to the rearrangement of the ligand
(Fig.\ \figClusterCoords). The timescale spectrum of C3 in Fig.\
\ref{fig:TA}a shows a broad distribution of times between 20 and
100\,ns, as well as a well-defined peak at $2\,\mu$s. This indicates
that the initial excitation of the $\alpha_3$-helix travels from
cluster C1 to cluster C3 either via rigid connections mediated, e.g.,
by cluster C2 and by the salt bridge Arg99-Glu34 (fast response), or
as a consequence of the 800\,ns conformational transition of C1 (slow
response). In the latter case, about half of the conformational
transitions of clusters C1 and C3 take place simultaneously (see Fig.\
\figCCtransitions), indicating possible cooperativity of these near-by
contacts.

%
%
\subsection{Long distance propagation}
\vspace*{-4mm}

The time evolution of the contact distances in Fig.\
\ref{fig:clusters}a indicate also a clear response of several clusters
that are distant to the initial photoexcitation of $\alpha_3$. This
holds to some extent for clusters C4 and C5 along the
$\alpha_2$-helix, and in particular for cluster C6, which we now focus
on. With 15 contacts involving mostly the loops $\beta_1$-$\beta_2$ and
$\beta_3$-$\alpha_1$, cluster C6 is clearly at the other end of
PDZ3. 
The time evolution of all contact distances of cluster C6 look very
similar, see Fig.\ \ref{fig:contacts} for $r_{18,-1}$ as a
representative example (and Fig.\ \figClusterCoords{} for all others).
They show an initial increase on a 200\,ns timescale, which peaks
around $1\,\mu$s and decays on a $2.5\,\mu$s timescale. Accordingly,
the timescale spectrum of C6 shows two well-defined peaks at 200\,ns
and $2.5\,\mu$s (Fig.\ \ref{fig:TA}a). Since on average no new
contacts are formed or broken (Fig.\ \ref{fig:clusters}d), and since
the final contact distances mostly return to their initial value,
cluster C6 appears to respond elastically. While this is true for most
contacts, however, a closer look shows that for $\sim 70\,\%$ of the
trajectories a new contact is formed between residues Arg18 and Gly83.

\begin{figure}[ht!]
	\centering
\includegraphics[width=0.35\textwidth]{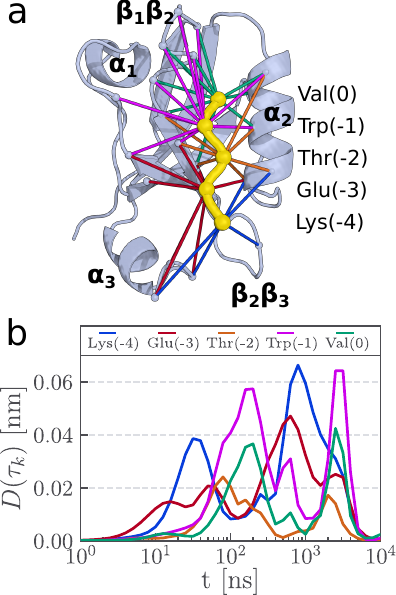}
\caption{
  (a) Inter-residue contacts of the ligand with PDZ3 and (b) their
  timescale spectrum.}
	\label{fig:ligand}
\end{figure}

To facilitate this response, cluster C6 need to be in some way
coupled to the initially perturbed cluster C1. For
example, there are two contacts of C1 that couple to the $\beta_2$
sheet (via residue Val28), which is connected to contacts of C6 involving
the $\beta_1$-$\beta_2$-loop (via residues Arg18 - Gly24). More importantly,
though, both C1 and C6 are well connected to the ligand, comprising 5
ligand contacts of C1 and 6 ligand contacts of C6.
To study how the ligand mediates the long-distant coupling between
clusters C1 and C6, we consider the contacts of the protein with the
five residues of the ligand. (See Fig.\ \figLigandContacts{} for the
time evolution of these contacts.) As shown in Fig.\
\ref{fig:ligand}a, Lys(-4) and Glu(-3) are close to cluster C1,
Thr(-2) is in the middle, and Trp(-1) and Val(0) are close to cluster
C6. Accordingly, cluster C1 contains 2 ligand contacts with Lys(-4)
and 3 with Glu(-3), while cluster C6 contains 4 ligand contacts with
Trp(-1) and 2 with Val(0).

Performing a timescale analysis of the contact distances associated
with the individual ligand residues, Fig.\ \ref{fig:ligand}b shows
that the main timescales of the contacts with Lys(-4) are very similar
to the ones of cluster C1 (i.e., 30 and 800\,ns), while the timescales
of the contacts with Val(0) are very similar to the ones of cluster C6
(i.e., 200\,ns and $2\,\mu$s). For the three inner ligand residues, we
find both timescales to gradually shift between these two limiting
cases. That is, the fast timescale (30\,ns for Lys(-4)) shifts to a
broad band covering $\sim 10$ to 100\,ns for Glu(-3), peaks around
100\,ns for Thr(-2), and arrives at 200\,ns for Trp(-1) and
Val(0). The slow timescale (800\,ns for Lys(-4)) evolves to a double
peak at $\sim 600$\,ns and $2\,\mu$s for Glu(-3), before it settles
around $2\,\mu$s for Trp(-1) and Val(0).

Hence the time-delayed response along the ligand indicates the
propagation of the perturbation of cluster C1 to the distant cluster
C6. In a first step, the detachment of $\alpha_3$ at 30\,ns appears to
cause a response of C6 at 200\,ns.  The associated propagation time of
$\sim$170\,ns corresponds to the time it takes to transfer the
structural strain of C1 via the ligand to C6. This is because this
transport represents a diffusive process that involves the
rearrangement of the conformation of the residues of the ligand and
the surrounding protein. In a second step, the alignment of $\alpha_3$
at 800\,ns pulls the ligand contacts of C1 again in the opposite
direction. This change also propagates through the ligand to cluster
C6, and facilitates the relaxation of the cluster distances to their
equilibrium value.

The response of the distant clusters is mainly elastic, that is,
hardly any new contacts are formed to stabilize the perturbed
structure. However, if the $\beta_1$-$\beta_2$- or
$\beta_3$-$\alpha_1$-loop of cluster C6 were involved in an
interaction with another nearby molecule (e.g., a protein or
co-factor, see Ref.\ \onlinecite{Ye13}), new contacts with this
molecule could stabilize the allosteric response of cluster C6,
resulting in an allosteric transition of C6 triggered by C1.
For example, the Rho GTPase Cdc42 was shown to allosterically regulate
Par-6 PDZ domain via binding to $\beta_1$ and
$\alpha_1$. \cite{Peterson04} 

Cluster C5 shares with C6 the same overall time evolution, i.e., an
transient response at 200\,ns, which decays on a timescale of
$2.5\,\mu$s, see Figs.\ \ref{fig:TA}a. Since C5 does not contain
ligand contacts, the signal transmission most likely proceeds via the
$\alpha_2$-helix and the four existing contacts with C6.  Finally,
clusters C4 and C7 show only a rather weak response to photoswitching
(Fig.\ \ref{fig:clusters}), and are therefore not considered here.


%
%
\section{Conclusions}
\vspace*{-4mm}

On the basis of extensive (in total $\sim 500\,\mu$s)
MD simulations and the identification of seven MoSAIC clusters of
highly correlated tertiary contacts, we have outlined a contact
cluster model of allosteric communication in PDZ3. Following an
external perturbation (such as photoswitching or the (un)binding of a
ligand), the protein experiences a structural strain that affects the
nearby contact clusters. The subsequent multi-step structural
reorganization process consists of cooperative conformational
transitions within a cluster (Fig.\ \ref{fig:contacts}), and of the
communication between clusters (Fig.\ \ref{fig:ligand}). The latter
represents a sequential process that is mediated by rigid secondary
structures. Hence we have shown that the response of the protein is in
general nonlinear (since contacts are broken and formed in the
clusters), and nonlocal (since rigid secondary structures connect
distant clusters).

This general picture of allostery is corroborated by the timescale
spectra of the contact clusters (Fig.\ \ref{fig:TA}) and the ligand
contacts (Fig.\ \ref{fig:ligand}), which illustrate the time evolution
of the multi-step process. We have found excellent overall agreement
between simulation and experiment,\cite{Bozovic21} which supports the
general idea that experimental and computational observables
essentially correspond to different projections of the underlying
transfer operator and therefore give similar timescales.\cite{Noe11}

It is instructive to contrast the contact cluster model with commonly
used network models,\cite{Guo16,Dokholyan16} which describe allosteric
communication in terms of an interaction matrix of the protein
residues. As an example of a quite elaborate approach, we mention the
work of Hamelberg and coworkers, \cite{Yao22} which is also based on
inter-residue contacts.  Performing MD simulations of the allosteric
states (e.g., active and inactive) of various systems, their
interaction matrix is constructed from the contact changes due to the
allosteric transition.  As in most network approaches, subsequently a
path analysis
is applied to the network, which yields the presumably most important
'allosteric pathways' from the source of the perturbation to the
target site.

Although our formulation bears some resemblance with this model
of contact changes, it is constructed from the structural evolution
revealed by the MD simulations, rather than from an empirical path
analysis. Network paths are given as a sequence of inter-residue
transitions, and as such cannot account for several main aspects
of the dynamical process underlying allostery. That is, the
cooperativity of contact changes (occurring in clusters), the
nonlocality of the (inter-cluster) interactions which couple remote
residues, and the possibility that several structural changes occur
simultaneously in a trajectory. While allosteric pathways of network
models may be a powerful tool, e.g., to predict the effects of
mutations, they do not necessarily reflect the time
evolution of a MD trajectory.

The discussion above indicates that the overall picture of
allostery mediated by communicating contact clusters may be
applicable to other allosteric systems. In particular, we have found
that the construction of contact clusters does not necessarily require
long MD data covering the allosteric transition, but can be also
achieved from standard equilibrium simulation of the two allosteric
states of the protein (e.g., the {\em cis} and {\em trans} states of
photoswitchable PDZ3, see Tab.\ \tabClusterCoords). To identify rigid
secondary structure elements that couple different clusters, a
rigidity analysis \cite{Hermans17} may be employed. The resulting
model may provide a qualitative description of possible allosteric
mechanisms for a given system. These insights may also be useful in
order to identify low-dimensional biasing coordinates, which can be
used by enhanced sampling techniques \cite{Mehdi24} that facilitate
the observation of allosteric transitions in large biomolecular systems.

%
%
\section{Methods}
 \vspace*{-4mm}

{\bf MD simulations.}
All simulations used the GROMACS v2020 software
package,\cite{Abraham15} the Amber99SB*ILDN force
field\cite{Best09} and the TIP3P water
model. As detailed in Ref.\ \onlinecite{Ali22}, we
first ran $8\!\times\! 10\,\mu$s-long MD simulations of the {\em cis}
and the {\em trans} equilibrium states of PDZ3. To study the time
evolution of the system following {\em cis} to {\em trans}
photoisomerization states of the azobenzene photoswitch, we sampled
100 statistically independent structures from the {\em cis}
equilibrium simulations, performed at time $t\!=\!0$ a
potential-energy surface switching method \cite{Nguyen06b} to mimic
the ultrafast photoisomerization of azobenzene, and calculated
$100 \!\times \!1\, \mu$s-long nonequilibrium trajectories, ten of
which were extended to $10\,\mu$s. Moreover, we chose new starting
structures from the {\em cis} equilibrium runs after 4 and 8\,$\mu$s,
and performed $12 \times 10\,\mu$s-long nonequilibrium simulations. 
All trajectories were sampled with a time step of 0.2\,ns, resulting in
total in $\sim 1.5\;10^6$ frames of the MD data.

{\bf MoSAIC analysis.}
First we identified the inter-residue contact of PDZ3 by assuming that a
contact is formed if the distance $r_{ij}$ between the closest
non-hydrogen atoms of residues $i$ and $j$ is shorter than
4.5\,\AA.\cite{Ernst15,Nagel23} Moreover, we requested
$|i-j| > 2 $ to neglect neighboring residues, and focused on contacts
that are populated more than $10\,\%$ of the simulation time. As
detailed in Ref.\ \onlinecite{Ali22}, this resulted in 403
inter-residue contacts.
To identify clusters of highly correlated contacts, we performed a
MoSAIC analysis\cite{Diez22} as explained above. To assess the
clearness of the definition of cluster, we repeated the MoSAIC analysis
for short ($1 \mu$s) nonequilibrium MD data \cite{Ali22} for long
nonequilibrium MD data, for {\em cis} and {\em trans} equilibrium
data, and for using the mutual information measure to calculate the
correlation. As shown in Tab.\ \tabClusterCoords\
listing the contacts of all clusters for all choices, this lead
to only minor changes of the clusters.

{\bf Timescale analysis.}
Using a maximum-entropy regularization method, \cite{Lorenz-Fonfria06}
we minimize $\chi^2-\lambda_\mathrm{reg}S_\mathrm{ent}$, with the
usual root mean square deviation $\chi^2$ of the fit function to the
data, the entropy-based regularization factor $S_\mathrm{ent}$, and
the regularization parameter chosen as $\lambda_\mathrm{reg}=100$. The
convergence of the fit is improved by extending the time series by an
additional time decade, with the extended value derived as average
over half of the previous decade.  To ensure an equal contribution of
each decade, we transformed the linearly spaced data into logarithmic
data with the same number of frames per decade.

%
%
\subsection*{Author's contributions}
\vspace*{-4mm}

All authors contributed equally to this work.

\subsection*{Supplementary material}
\vspace*{-4mm} 

Includes a table of the contacts of all clusters, and figures showing the
time evolution of all all considered contact distances.

\subsection*{Acknowledgments}
\vspace*{-4mm}

The authors thank Peter Hamm, Steffen Wolf, Georg Diez, Daniel Nagel
and Sofia Sartore for helpful comments and discussions. This work has
been supported by the Deutsche Forschungsgemeinschaft (DFG) within the
framework of the Research Unit FOR 5099 ``Reducing complexity of
nonequilibrium'' (project No.~431945604), the High Performance and
Cloud Computing Group at the Zentrum f\"ur Datenverarbeitung of the
University of T\"ubingen, the state of Baden-W\"urttemberg through
bwHPC and the DFG through grant no INST 37/935-1 FUGG (RV bw16I016),
the Black Forest Grid Initiative, and the Freiburg Institute for
Advanced Studies (FRIAS) of the Albert-Ludwigs-University Freiburg.

\subsection*{Data Availability Statement}
\vspace*{-4mm}

The clustering package MoSAIC is available at our group homepage 
\url{https://www.moldyn.uni-freiburg.de/software.html}. Trajectories
and simulation structures are available from the authors upon 
reasonable request.

%
%
\bibliography{new}

\end{document}


\setstretch{1}

\begin{figure}[h!]
	\centering
	{\includegraphics[width=0.9\textwidth]{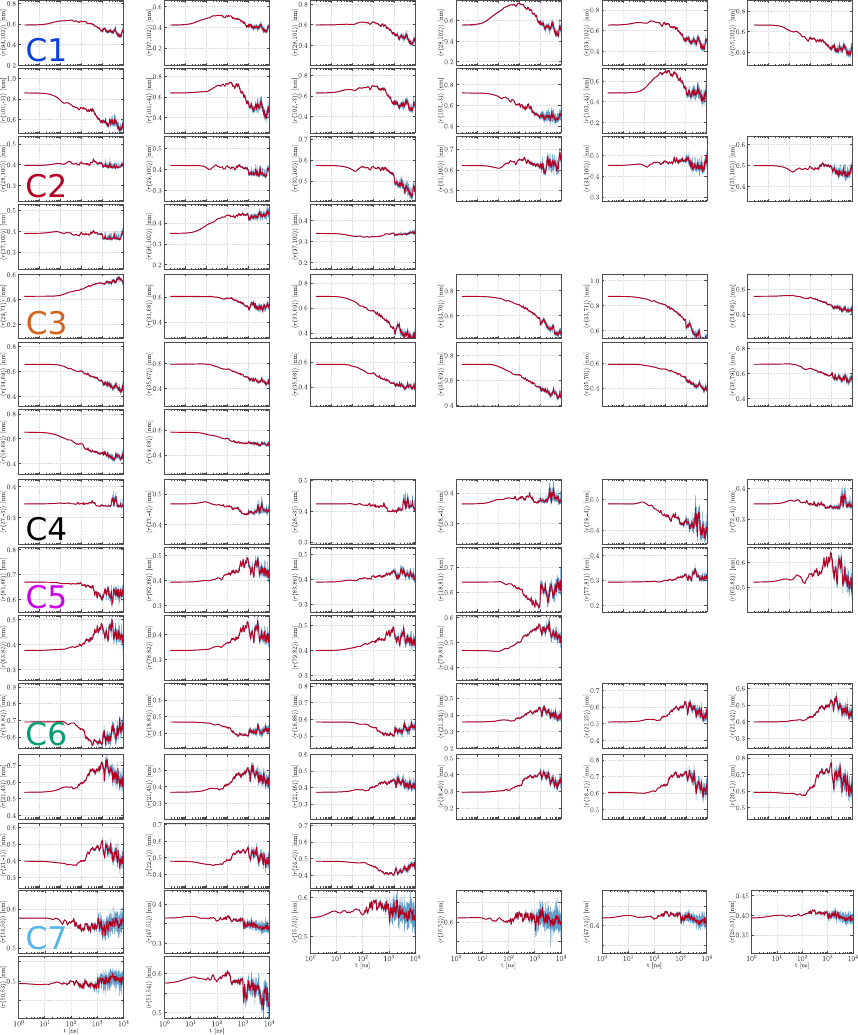}}
	\caption{Averaged contact distances $r_{ij}(t)$ for all 7 clusters, using a logarithmic Gaussian filter of $\sigma\!=\!6$ frames. The y-axis is chosen for each cluster such that it covers the same range around the initial position $r_{ij}(t\!=\!0)$. The blue line reflects the fluctuation from the unfiltered data while the red line correspond to the filtered data with $\sigma\!=\!6$ frames.}
	\label{fig1}
\end{figure}

\begin{figure}[h!]
\centering
{\includegraphics[width=0.45\textwidth]{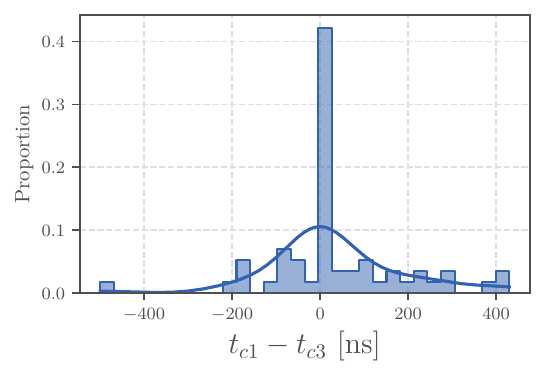}}
\caption{Histogram of the time difference ($t_{C1}\!-\!t_{C3}$) between transition times in single 1\,$\mu$s nonequilibrium trajectories in clusters C1 and C3 .The transition time is defined as the time required for a contact distance to change its status from high amplitude phase to contact phase (or vice versa) after the photoswitching event at $t\!=\!0$. }
\label{fig2}
\end{figure}

\begin{figure}[h!]
	\centering
	{\includegraphics[width=0.95\textwidth]{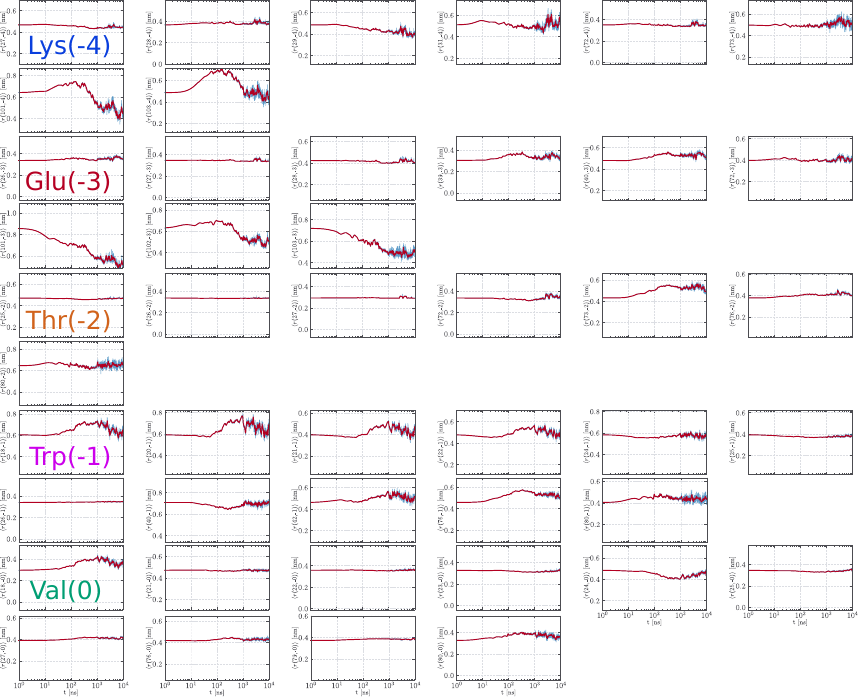}}
	\caption{Averaged contact distances $r_{ij}(t)$ for all contacts to the 5 ligand residues (total of 45 distances). A logarithmic Gaussian filter of $\sigma\!=\!6$ frames is used, and as above, the y-axis is set for all distances such that it covers the same range around the initial position.}
	\label{fig3}
\end{figure}


\clearpage
\section*{Compaison of MoSAIC clustering for different data sets}
The following data sets are compared:\\
\textbf{`Short NEQ'} contains $100 \!\times\! 1\, \mu$s nonequilibrium trajectories. The contacts (in total 403) of this clustering (with Leiden resolution parameter $\gamma\!=\!0.5$) is shown in Fig.\,1 in the main text. In Table.\,S1, the less important clusters C2, C4, C5 and C7 are only given for this data set, see below. For the most relevant clusters C1, C3 and C6, we also show the contact distances of three other data sets.

\textbf{`Long NEQ'} contains $22 \!\times\! 10\, \mu$s nonequilibrium trajectories with Leiden resolution parameter $\gamma \!= \!0.3$.

\textbf{`EQ'} contains $0.5 \!\times\! 10^6$ MD frames for each \textit{Cis} and \textit{Trans} equilibrium simulations with $\gamma \!= \!0.4$.

\textbf{`Short NEQ/MI'} employ (instead of the linear correlation used above) the mutual information as correlation measure with $\gamma \!= \!0.03$.

\clearpage	
\renewcommand*{\arraystretch}{1.3}
\begin{table}[ht!]
	\label{table:clusters}
	\caption{List of contacts distances $r_{ij}$ between residue $i$ and $j$ of Leiden clusters C1-C8. The result of four data sets are compared for C1, C3 and C6.}
	\begin{tabular}{cl}
		\hline
		Cluster & Coordinates \\
		\hline 
		1&\\
		Short NEQ& $r_{94, 102}$, $r_{97, 102}$, $r_{28, 101}$, $r_{28, 102}$, $r_{39, 102}$, $r_{55, 102}$, $r_{101, (-3)}$, $r_{101, (-4)}$, $r_{102, (-3)}$,\\& $r_{103, (-3)}$, $r_{103, (-4)}$\\
		Long NEQ& $r_{94, 98}$, $r_{94, 102}$, $r_{95, 99}$, $r_{97, 102}$, $r_{28, 101}$, $r_{28, 102}$, $r_{39, 97}$, $r_{39, 102}$, $r_{55, 102}$, $r_{101, (-3)}$, \\& $r_{101, (-4)}$, $r_{102, (-3)}$, $r_{102, (-4)}$, $r_{103, (-3)}$, $r_{103, (-4)}$\\ 
		EQ& $r_{94, 102}$, $r_{97, 102}$, $r_{28, 101}$, $r_{28, 102}$, $r_{39, 102}$, $r_{101, (-3)}$, $r_{101, (-4)}$, $r_{102, (-3)}$, $r_{103, (-3)}$, \\&$r_{103, (-4)}$\\
		Short NEQ\textbackslash MI& $r_{93, 97}$, $r_{94, 98}$, $r_{94, 102}$, $r_{95, 98}$, $r_{95, 99}$, $r_{97, 102}$, $r_{28, 101}$, $r_{28, 102}$, $r_{39, 97}$, $r_{39, 102}$, $r_{55, 102}$,\\& $r_{101, (-3)}$, $r_{101, (-4)}$, $r_{102, (-3)}$, $r_{103, (-3)}$, $r_{103, (-4)}$ \\
		\hline
		3&\\
		Short NEQ& $r_{29, 71}$, $r_{33, 68}$, $r_{33, 69}$, $r_{33, 70}$, $r_{33, 71}$, $r_{34, 68}$, $r_{34, 69}$, $r_{35, 67}$, $r_{35, 68}$, $r_{35, 69}$, $r_{35, 70}$,  $r_{36, 79}$, \\&  $r_{58, 68}$,  $r_{59, 68}$  \\
		Long NEQ&$r_{30, 36}$, $r_{30, 72}$, $r_{32, 69}$, $r_{32, 70}$, $r_{32, 72}$, $r_{33, 69}$, $r_{33, 70}$, $r_{33, 71}$, $r_{34, 68}$, $r_{34, 69}$, $r_{34, 70}$, $r_{35, 67}$, \\& $r_{35, 68}$, $r_{35, 69}$, $r_{35, 70}$, $r_{36, 79}$, $r_{58, 68}$, $r_{59, 68}$, $r_{60,68}$, $r_{61, 68}$, $r_{62, 86}$, $r_{67, 79}$\\
		EQ& $r_{30,36}$, $r_{32,69}$, $r_{32,70}$, $r_{32,71}$, $r_{33, 69}$, $r_{33, 70}$, $r_{33, 71}$, $r_{34, 68}$, $r_{34, 69}$, $r_{34,70}$, $r_{35, 68}$, $r_{35, 69}$, \\& $r_{35,70}$,$r_{35, 70}$,  $r_{58, 68}$, $r_{59, 68}$\\ 
		Short NEQ\textbackslash MI & $r_{28, 72}$, $r_{29, 71}$, $r_{30, 72}$, $r_{32, 69}$, $r_{32, 70}$, $r_{32, 71}$, $r_{33, 68}$, $r_{33, 69}$, $r_{33, 70}$, $r_{33, 71}$, $r_{34, 68}$, $r_{34, 69}$, \\& $r_{34, 70}$, $r_{35, 67}$, $r_{35, 68}$, $r_{35, 69}$, $r_{35, 70}$,  $r_{36, 79}$, $r_{58, 68}$,  $r_{59, 68}$, $r_{73, (-2)}$\\
		\hline
		6&\\
		Short NEQ& $r_{18, 82}$, $r_{18, 83}$, $r_{18, 86}$, $r_{21, 24}$, $r_{21, 25}$, $r_{21, 42}$, $r_{21, 43}$, $r_{21, 45}$, $r_{21, 46}$, $r_{18, (0)}$, $r_{18, (-1)}$,  \\& $r_{20, (-1)}$, $r_{21, (-1)}$, $r_{22, (-1)}$, $r_{24, (0)}$ \\
		Long NEQ& $r_{17, 22}$, $r_{18, 21}$, $r_{18, 22}$, $r_{18, 23}$, $r_{19, 22}$, $r_{21, 24}$, $r_{21, 25}$, $r_{17, 46}$ $r_{18, 80}$, $r_{19, 46}$, $r_{19, 84}$,$r_{21, 42}$,\\& $r_{21, 43}$, $r_{21, 45}$, $r_{21, 46}$, $r_{24, 45}$ $r_{24, 46}$, $r_{25, 42}$, $r_{26, 42}$, $r_{18, (0)}$, $r_{18, (-1)}$, $r_{20, (-1)}$, $r_{21, (-1)}$, \\& $r_{22, (-1)}$, $r_{24, (0)}$, $r_{42, (-1)}$\\
		EQ& $r_{18,21}$, $r_{19,22}$, $r_{21, 24}$, $r_{21, 25}$, $r_{19,84}$, $r_{21, 42}$, $r_{21, 43}$, $r_{21, 45}$, $r_{21, 46}$, $r_{18, (0)}$, $r_{18, (-1)}$, \\& $r_{20, (-1)}$, $r_{21, (-1)}$, $r_{22, (-1)}$, $r_{24, (0)}$, $r_{42,(-1)}$\\
		Short NEQ\textbackslash MI & $r_{18, 21}$, $r_{18, 82}$, $r_{18, 83}$, $r_{18, 86}$, $r_{19, 22}$, $r_{21, 24}$, $r_{21, 25}$, $r_{21, 42}$, $r_{21, 43}$, $r_{21, 45}$, $r_{21, 46}$, $r_{24, 46}$, \\& $r_{18, (0)}$, $r_{18, (-1)}$, $r_{20, (-1)}$, $r_{21, (-1)}$, $r_{22, (-1)}$, $r_{23, (0)}$, $r_{24, (0)}$, $r_{24, (-1)}$, $r_{25, (0)}$,$r_{42, (-1)}$\\
		\hline
	\end{tabular}
\end{table}
\begin{table}[ht!]
	\begin{tabular}{cl}
		\hline
		Cluster & Coordinates \\
		\hline
		2& $r_{28, 100}$, $r_{29, 100}$, $r_{30, 100}$, $r_{31, 100}$, $r_{34, 100}$, $r_{35, 100}$, $r_{37, 100}$, $r_{96, 100}$, $r_{97, 100}$ \\
		4& $r_{27, (-3)}$, $r_{27, (-4)}$, $r_{28, (-3)}$, $r_{28, (-4)}$,  $r_{29, (-4)}$,   $r_{72, (-4)}$\\
		5& $r_{81, 86}$, $r_{82, 86}$, $r_{83, 86}$, $r_{18, 81}$, $r_{77, 81}$, $r_{62, 82}$, $r_{63, 82}$, $r_{78, 82}$, $r_{79, 82}$,    $r_{79, 83}$  \\
		7& $r_{47, 51}$, $r_{47, 52}$, $r_{14, 50}$, $r_{50, 53}$, $r_{51, 54}$, $r_{16, 52}$, $r_{15, 52}$, $r_{25, 53}$ \\
		8& $r_{1, 95}$, $r_{2, 95}$, $r_{1, 98}$, $r_{2, 51}$, $r_{1, 51}$, $r_{1, 102}$, $r_{2, 98}$, $r_{2, 52}$, $r_{3, 54}$, $r_{2, 102}$, $r_{1, 100}$, $r_{1, 48}$, $r_{3, 95}$, $r_{2, 94}$,\\& $r_{2, 53}$, $r_{2, 54}$, $r_{2, 97}$\\
		\hline
	\end{tabular}
\end{table}